\documentclass{elsarticle}

\usepackage{amsmath, amssymb}

\usepackage{graphicx}
\graphicspath{%
    {converted_graphics/}
    {/}
}
\begin{document}

\title{Some Models for Epidemics of Vector-Transmitted Diseases}

\author{Fred Brauer \footnote{University of British Columbia, Vancouver BC, Canada, corresponding author (brauer@math.ubc.ca)}
Carlos Castillo-Chavez \footnote{Arizona State University, Tempe  AZ, USA}
Anuj Mubayi \footnote{Arizona State University, Tempe  AZ, USA}
Sherry Towers \footnote{Arizona State University, Tempe  AZ, USA}}     

\date{\today}          
\maketitle

Abstract: Vector-transmitted diseases such as dengue fever and chikungunya have been spreading rapidly in many parts of the world.
The Zika virus has been known since 1947 and invaded South America in 2013. It can be transmitted not only by  (mosquito) vectors but also directly through sexual contact. Zika has developed into a serious global health problem 
because, while most cases are asymptomatic or very light, babies born to Zika - infected mothers may develop microcephaly and other very serious birth defects.

We formulate and analyze two epidemic models for vector-transmitted diseases, one appropriate for dengue and chikungunya fever outbreaks and one that includes direct transmission appropriate for Zika virus outbreaks. This is especially important because the Zika virus is the first example of a disease that can be spread both indirectly through a vector and directly (through sexual contact).  In both cases, we obtain expressions for the basic reproduction number and show how to use the initial exponential growth rate to estimate the basic reproduction number. However, for the model that includes direct transmission some additional data would be needed to identify the fraction of cases transmitted directly. Data for the 2015 Zika virus outbreak in Barranquilla, Colombia has been used to fit parameters to the model developed here and to estimate the basic reproduction number.

\maketitle

\section{Introduction}

The demonstration in 1897 by Dr. R.A. Ross that malaria is transmitted from person to person through a vector, the Anopheles mosquito, was a landmark in the early history of mathematical epidemiology; for this achievement Dr. Ross was awarded the 1902 Nobel Prize in Medicine. Malaria is endemic, with 90 \% of its cases in Sub-Saharan Africa and causes hundreds of thousands of deaths annually, mostly children less than five years old.

While malaria is one of the communicable diseases with the largest number of cases and deaths in the world, there are other diseases transmitted by vectors of great importance including dengue fever and, more recently chikungunya and Zika virus. These diseases often appear as epidemic outbreaks rather than as endemic diseases. The purpose of this paper is to describe and analyze some epidemic models for diseases which are transmitted by vectors.

While there have been cases of probable dengue fever more than 1000 years ago, the first recognized dengue epidemics occurred in Asia, Africa, and North America in the 1780's. There have been frequent outbreaks since then, and the number of reported cases has been increasing rapidly recently. According to the World Health Organization, approximately 50,000,000 people worldwide are infected with dengue. Symptoms may include fever, headaches, joint and muscle pain, and nausea, but many cases are very mild. There is no cure for dengue fever, but most patients recover with rest and fluids. There are at least four different strains of dengue fever, and there is some cross-immunity between strains. Dengue fever is transmitted by the mosquito {\it Aedes aegypti}, and most control strategies are aimed at mosquito control.

Another disease transmitted by vectors, in fact the same {\it Aedes aegypti} mosquito that transmits dengue, is the Zika virus. The Zika virus was first observed about 1952, but initially cases were rare. In 2007 a major epidemic occurred in Yap Island, Micronesia.
Since April, 2015 there has been a large continuing outbreak of Zika virus that started in Brazil and has spread to much of South and Central America. It has become a major concern because it is now established that there is some correlation with very serious birth defects in babies born to infected mothers \cite{S-F16}. A new feature of the Zika virus that has been identified is that infection may be transmitted directly by blood transfusions and sexual contact \cite{Metal15} as well as through vectors.

In this paper we will formulate and analyze two epidemic models. The first is a general vector transmission model of $SEIR$ type for human hosts and of $SEI$ type for vectors. It does not capture all aspects of dengue fever, because there are at least four  strains of dengue, with some cross-immunity between strains, but can be viewed as a general starting point for the study of vector models. The second model, can be viewed as a basic Zika virus model; it adds direct transmission of infection to the first model.
It would be possible to extend this model further by separating human hosts into males and females and treating direct disease transmission as sexual transmission. 

The work described here grew out of a study of the 2015 Zika outbreak in Barranquilla, Colombia \cite{Setal16}. We are indebted to Dr. Andrew Falconer and Dr. Claudia Romero-Vivas for some very useful comments on models and data. We are also indebted to Jim Cushing and Pauline van den Driessche for some new insights into basic reproduction numbers and the next generation matrix.

\section{A Basic Vector Transmission Model} \label{first}

We begin with a basic epidemic model for a vector-transmitted disease. We will assume throughout that vectors satisfy a simple $SEI$ model, with no recovery from infection. We are thinking of mosquitoes as vectors, and because a mosquito lifetime is much shorter than that of the human hosts we must include demographics in the vector population. 

We consider a constant total population size $N$ of hosts (humans), divided into $S$ susceptibles, $E$ exposed members, $I$ infectives, and $R$ removed members. A host makes an average of $\beta$ contacts sufficient to receive infection in unit time from vectors. The contact rate $\beta$ is a product of two factors, namely the biting rate $a$ and the probability $f_{vh}$ that a bite transmits infection from vector to human,
\[
\beta = a f_{vh}.
\] 
Exposed hosts proceed to the infective class at rate $\kappa$ and infected hosts recover at rate $\alpha$. The total number of contacts by humans sufficient to transmit infection is $\beta N$.

The number of vectors (mosquitoes) is $N_v$ divided into $S_v$ susceptibles, $E_v$ exposed members, and $I_v$ infectives. Each vector makes $\beta_v$ contacts sufficient to receive infection from human hosts in unit time. The contact rate $\beta_v$ is a product of two factors, namely the biting rate $a_v$ and the probability $f_{hv}$ that a bite transmits infection from human to vector,
\[
\beta_v = a_v f_{hv}.
\] 
There is a constant birth rate $N_v$ of vectors in unit time and a proportional vector death rate $\mu$ in each class, so that the total vector population size $N_v$ is constant. Exposed vectors move to the infected class at rate $\eta$ and do not recover from infection. The total number of contacts
by vectors sufficient to transmit infection is $\beta_v N_v$. 

Because the total number of bites received by humans must be equal to the total number of bites made by mosquitoes, we have a balance relation
\[
a N = a_v N_v.
\]
or
\begin{equation} \label{bal}
f_{hv} \beta N = f_{vh} \beta_v N_v.
\end{equation}
This balance relation must hold at every time $t$. We think of $a_v$, the size of a mosquito blood meal, and $f_{hv}$ as fixed. Thus $\beta_v$ is also fixed. The number of effective bites of a human in unit time is
\[
\beta = \beta_v \frac{f_{vh}}{f_{hv}}\frac{N_v}{N}.
\]
We are assuming that the population sizes $N$ and $N_v$ are constant, but it is important to remember that if one of the population sizes changes, for example because of a program to kill mosquitoes, a change in the value of $\beta$ would be a consequence. 

A susceptible human receives $\beta$ effective mosquito bites in unit time, of which a fraction $I_v/N_v$ is with an infective mosquito. Thus the number of new infective humans in unit time is
\[
\beta S \frac{I_v}{N_v}.,
\]
which, because of \eqref{bal}, is equal to
\[
\beta_v S \frac{I_v}{N}.
\]
A similar argument shows that the number of new mosquito infections is
\[
\beta_v S_v \frac{I}{N}.
\]
The model is
\begin{eqnarray} \label{model}
S' &=& - \beta S \frac{I_v}{N_v}   \\
E' &=&  \beta S \frac{I_v}{N_v}  - \kappa E \nonumber \\
I' &=& \kappa E - \gamma I \nonumber \\
S_v' &=& \mu N_v - \mu S_v - \beta_v S_v \frac{I}{N} \nonumber \\
E_v' &=&  \beta_v S_v \frac{I}{N} - (\mu + \eta) E_v \nonumber \\
I_v' &=& \eta E_v - \mu I_v \nonumber
\end{eqnarray}

\subsection{The basic reproduction number}

The basic reproduction number is defined as the number of secondary disease cases caused by introducing a single infective human into a wholly susceptible population of both hosts (humans)  and vectors (mosquitoes). For the model \eqref{model} this may be calculated directly. There are two stages. First, the infective human infects mosquitoes, at a rate
$\beta_v N/N_v$ for a time $1/\gamma$. This produces $\beta_v N/N_v \gamma$ infected mosquitoes, of whom a fraction
$\eta/(\eta + \mu)$ proceeds to become infectious.

The second stage is that these infective mosquitoes infect humans at a rate $\beta N_v/N$ for a time $1/\mu$, producing
$\beta N_v/N \mu$ infected humans per mosquito. The net result of these two stages is
\[
\frac{\beta_v N}{N_v \gamma}\frac{\eta}{\eta + \mu}\frac{\beta N_v}{N \mu}  = \beta_v \beta \frac{\eta}{(\eta + \mu)\gamma \mu}
\]
infected humans, and this is the basic reproduction number $\mathcal{R}_0$. 

We could also calculate the basic reproduction number by using the next generation matrix approach \cite{vdDW02}. This would give the next generation matrix
\[
K 
\begin{bmatrix}
0  & \beta\frac{N}{N_v}\frac{\eta}{\mu(\mu + \eta)}  \\
 \beta_v\frac{N_v}{N}\frac{1}{\gamma} & 0 
\end{bmatrix}.
\]  
The basic reproduction number is the positive eigenvalue of this matrix, 
\[
\mathcal{R}_0 = \sqrt{\beta \beta_v \frac{\eta}{\mu \gamma(\mu + \eta)}}.
\]

In this calculation, the transition from host to vector to host is considered as two generations. In studying vector transmitted diseases it is common to consider this as one generations and use the value that we obtained by our direct approach
\begin{equation} \label{repno1}
\mathcal{R}_0 = \beta \beta_v \frac{\eta}{\mu \gamma(\mu + \eta)}.
\end{equation}
This choice is made in \cite{Cetal07} and \cite{Ketal16} and is the choice that we make because it conforms to the result obtained directly. . However, other references, including 
\cite{Petal10}, use the square root form, and it is important to be aware of which form is being used in any study. The two choices have the same threshold value. 

\subsection{The initial exponential growth rate}

In order to determine the initial exponential growth rate from the model, a quantity that can be compared with experimental data, we linearize the model \eqref{model} about the disease-free equilibrium $S = N, E = I = 0, S_v = N_v, E_V = I_v = 0.$ If we let
$y= N - S, z = N_v - S_v$, we obtain the linearization
\begin{eqnarray} \label{lin}
y' &=& \beta N \frac{I_v}{N_v}   \\
E' &=&  \beta N \frac{I_v}{N_v}  - \kappa E \nonumber \\
I' &=& \kappa E - \gamma I \nonumber \\
z' &=& - \mu z + \beta_v N_v \frac{I}{N} \nonumber \\
E_v &=&  \beta_v N_v \frac{I}{N} - (\mu + \eta) E_v \nonumber \\
I_v' &=& \eta E_v - \mu I_v \nonumber
\end{eqnarray}
The corresponding characteristic equation is
\[
det \begin{bmatrix}
- \lambda & 0 & 0 & 0& 0 & \beta \frac{N}{N_v} \\
0 & - (\lambda + \kappa) & 0 & 0 & 0 & \beta \frac{N}{N_v} \\
0 & \kappa & - (\lambda + \gamma) & 0 & 0 & 0 \\
0 & 0 & - \beta_v \frac{N_v}{N} & - (\lambda + \mu) & 0 & 0 \\
0 & 0 & - \beta_v \frac{N_v}{N} & 0 & -  (\lambda + \mu + \eta) & 0  \\
0 & 0 & 0 & 0 & \eta & - (\lambda + \mu) 
\end{bmatrix}
= 0.
\]
Solutions of the linearization \eqref{lin} are linear combinations of exponentials whose exponents are the roots of the characteristic equation (the eigenvalues of the coefficient matrix of \eqref{lin}).

We can reduce this equation to a product of two factors and a fourth degree polynomial equation
\[
\lambda(\lambda + \mu) 
det \begin{bmatrix}
-(\lambda + \kappa) & 0 & 0 & \beta\frac{N}{N_v} \\
\kappa & - (\lambda + \gamma) & 0 & 0 \\
0 & \beta_v \frac{N_v}{N} & - (\lambda + \mu + \eta) & 0 \\
0 & 0 & \eta & - (\lambda + \mu) 
\end{bmatrix}
= 0.
\]

The initial exponential growth rate is the largest root of this fourth degree equation, which reduces to
\begin{equation} \label{grow}
g(\lambda) = (\lambda + \kappa)(\lambda + \gamma)(\lambda + \mu + \eta)(\lambda + \mu) - \beta \beta_v \kappa \eta = 0.
\end{equation} 
Since $g(0) < 0$ if $\mathcal{R}_0 > 1$, and since $g(\lambda)$ is positive for large positive $\lambda$ and $g'(\lambda) > 0$ for positive 
$\lambda$, there is a unique positive root of the equation $g(\lambda) = 0$, and this is the initial exponential growth rate. The initial exponential growth rate may be measured experimentally, and if the measured value is $\rho$, then from \eqref{grow} we obtain
\[
(\rho + \kappa)(\rho + \gamma)(\rho + \mu + \eta)(\rho + \mu) =  \beta \beta_v \kappa \eta = \mathcal{R}_0 \kappa  \mu \gamma(\mu + \eta).
\]
from this, we obtain
\begin{equation} \label{rep}
\mathcal{R}_0 = \frac{(\rho + \kappa)(\rho + \gamma)(\rho + \mu + \eta)(\rho + \mu)}{\kappa \mu \gamma (\mu + \eta)}
\end{equation}
This gives a way to estimate the basic reproduction number from measurable quantities. The question of estimating parameters for the Zika virus outbreak of 2015 has been described extensively in \cite{Setal16}. In addition, the balance relation 
\eqref{bal} allows us to calculate the values of $\beta$ and $\beta_v$ separately, which makes it possible to simulate the model. This is essential since there does not appear to be a final size relation for the model \eqref{model}.

\section{Inclusion of direct transmission}

For the Zika virus, it has been established that in addition to vector transmission of infection there may also be direct transmission through sexual contact. The Zika virus is the first example of an infection that can be transferred both directly and through a vector, and it is important to include direct transmission (in this case sexual transmission) in a model. To model this, we add to the model \eqref{model} a term $\alpha S \frac{I}{N}$ describing a rate $\alpha$ of movement from $S$ to $E$. This leads to the model
\begin{eqnarray} \label{model2}
S' &=& - \beta S \frac{I_v}{N_v} - \alpha S \frac{I}{N}  \\
E' &=&  \beta S \frac{I_v}{N_v} + \alpha S \frac{I}{N} - \kappa E \nonumber \\
I' &=& \kappa E - \gamma I \nonumber \\
S_v' &=& \mu N_v - \mu S_v - \beta_v S_v \frac{I}{N} \nonumber \\
E_v' &=&  \beta_v S_v \frac{I}{N} - (\mu + \eta) E_v \nonumber \\
I_v' &=& \eta E_v - \mu I_v \nonumber
\end{eqnarray}

The rate $\alpha$ is an average over the human population; if transmission is possible only from male to female this is incorporated into $\alpha$.

To calculate the basic reproduction number $\mathcal{R}_0$, we use the same direct approach as that used in Section \ref{first}.
If there is sexual transmission, this operates independent of the host-vector interaction, and produces $\alpha$ cases in unit time for a time $1/\gamma$, adding a simple term $\alpha/\gamma$ to the reproduction number
\begin{equation} \label{repno2}
\mathcal{R}_0 = \beta \beta_v \frac{\eta}{\mu \gamma(\mu + \eta)} + \frac{\alpha}{\gamma}.
\end{equation}

We define
\[
\mathcal{R}_v = \beta \beta_v \frac{\eta}{\mu \gamma(\mu + \eta)},
\]
the vector transmission reproduction number, and 
\[
\mathcal{R}_d = \frac{\alpha}{\gamma},
\]
the direct transmission reproduction number, so that
\[
\mathcal{R}_0 = \mathcal{R}_v + \mathcal{R}_d.
\]

If we use the next generation matrix approach, using the same approach as that used in Section \ref{first},
we form the matrix product
$K_L = F V^{-1}$ with 
\[
F =
\begin{bmatrix}
0 & \alpha                     & 0 & \beta\frac{N}{N_v} \\
0 & 0                     & 0 & 0 \\
0 & \beta_v \frac{N_v}{N} & 0 & 0 \\
0 & 0                     & 0 & 0 
\end{bmatrix} 
\quad V =
\begin{bmatrix}
\kappa  & 0      & 0         & 0 \\
-\kappa & \gamma & 0         & 0 \\
0       & 0      & \mu +\eta & 0 \\
0       & 0      & - \eta    & \mu 
\end{bmatrix}
\]
Then the next generation matrix with large domain is
\[
K_L =
\begin{bmatrix}
\frac{\alpha}{\gamma} & \frac{\alpha}{\gamma} & \beta\frac{N}{N_v}\frac{\eta}{\mu(\mu + \eta)}   & \beta\frac{N}{N_v}\frac{1}{\mu} \\
0 & 0 & 0 & 0 \\
\beta_v\frac{N_v}{N}\frac{1}{\gamma} & \beta_v\frac{N_v}{N}\frac{1}{\gamma} & 0 & 0 \\
0 & 0 & 0 & 0
\end{bmatrix} 
\]

 The next generation matrix $K$ is the $2 \times 2 $ matrix
\[
K =
\begin{bmatrix}
\frac{\alpha}{\gamma}  & \beta\frac{N}{N_v}\frac{\eta}{\mu(\mu + \eta)}  \\
 \beta_v\frac{N_v}{N}\frac{1}{\gamma} & 0 
\end{bmatrix}.
\]  

The positive eigenvalue of this matrix is
\begin{eqnarray*}
\lambda &=& \frac{\alpha}{2\gamma} + \frac{1}{2} \sqrt{\frac{\alpha^2}{\gamma^2} + 4 \mathcal{R}_v} \\
        &=& \frac{1}{2}\left[\mathcal{R}_d + \sqrt{\mathcal{R}_d^2 + 4\mathcal{R}_v}\right].
\end{eqnarray*}

We may calculate that $\lambda = 1$ if and only if
\[
 \mathcal{R}_v + \mathcal{R}_d = 1.
\]

We now have two potential expressions for the basic reproduction number, namely $\mathcal{R}_v + \mathcal{R}_d$ and 
\[
\mathcal{R}^* = \frac{1}{2}[\mathcal{R}_d + \sqrt{\mathcal{R}_d^2 + 4\mathcal{R}_v}].
\]

Different expressions are possible for the next generation matrix and that these may lead to different expressions for the basic reproduction number. This is shown in \cite{CD16}, and was drawn to our attention by Jim Cushing.  

The expression $\mathcal{R}_v + \mathcal{R}_d$ appears to us to be a more 
natural form than $\mathcal{R}^*$, and we choose to use this for the basic reproduction number. It can be obtained from the following expression for the next generation matrix, suggested to us by Pauline van den Driessche. We consider only human infections as new infections, and take

\[
F =
\begin{bmatrix}
0 & \alpha                     & 0 & \beta\frac{N}{N_v} \\
0 & 0                     & 0 & 0 \\
0 & 0 & 0 & 0 \\
0 & 0                     & 0 & 0 
\end{bmatrix} 
\quad V =
\begin{bmatrix}
\kappa  & 0      & 0         & 0 \\
-\kappa & \gamma & 0         & 0 \\
0       & - \beta_v \frac{N_v}{N}      & \mu +\eta & 0 \\
0       & 0      & - \eta    & \mu 
\end{bmatrix}
\]

Then 
\[
V^{-1} = 
\begin{bmatrix}
\frac{1}{\kappa} & 0 & 0 & 0 \\
\frac{1}{\gamma} & \frac{1}{\gamma} &0 & 0 \\
\beta_v\frac{N_v}{N} \frac{1}{\gamma(\mu + \eta)} & \beta_v\frac{N_v}{N} \frac{\eta}{\gamma(\mu + \eta)} & \frac{1}{\mu + \eta} & 0 \\
\beta_v\frac{N_v}{N} \frac{\eta}{\mu \gamma(\mu + \eta)} &  \beta_v\frac{N_v}{N} \frac{\eta}{\mu \gamma(\mu + \eta)}                                                        & \frac{\eta}{\mu(\mu + \eta)} & \frac{1}{\mu} 
\end{bmatrix}
\]

Since only the first row of $F$ has non-zero entries, the same is true of $FV^{-1}$, and from this we can deduce that the only non-zero eigenvalue of $FV^{-1}$ is the entry in the first row, first column of $FV^{-1}$, and this is
\[
\frac{\alpha}{\gamma} + \beta \beta_v \frac{\eta}{\mu \gamma (\mu + \eta)} = \mathcal{R}_d + \mathcal{R}_v = \mathcal{R}_0.
\]

If we use this somewhat unorthodox approach to the next generation matrix for the model \eqref{model}, we obtain the form 
$\mathcal{R}_v$, with no square root, for the reproduction number. We now have two viable expressions for the basic reproduction number, namely $\mathcal{R}^*$ and $\mathcal{R}_0$, both derived from a next generation matrix approach but with different separations. We have chosen to use $\mathcal{R}_0$ for the basic reproduction number because it is more readily interpreted as a number of secondary infections. Other sources, including \cite{Getal16}, use $\mathcal{R}^*$. In studying data for epidemic models that include vector transmission it is absolutely vital to specify exactly what form is being used for the basic reproduction number.

\subsection{The initial exponential growth rate}

In order to determine the initial exponential growth rate from the model, a quantity that can be compared with experimental data, we linearize the model \eqref{model} about the disease-free equilibrium $S = N, E = I = 0, S_v = N_v, E_V = I_v = 0.$ If we let
$y= N - S, z = N_v - S_v$, we obtain the linearization
\begin{eqnarray} \label{lin}
y' &=& \beta N \frac{I_v}{N_v}  + \alpha I \\
E' &=&  \beta N \frac{I_v}{N_v} + \alpha I - \kappa E \nonumber \\
I' &=& \kappa E - \gamma I \nonumber \\
z' &=& - \mu z + \beta_v N_v \frac{I}{N} \nonumber \\
E_v &=&  \beta_v N_v \frac{I}{N} - (\mu + \eta) E_v \nonumber \\
I_v' &=& \eta E_v - \mu I_v \nonumber
\end{eqnarray}
The corresponding characteristic equation is
\[
det \begin{bmatrix}
- \lambda & 0 & \alpha & 0& 0 & \beta \frac{N}{N_v} \\
0 & - (\lambda + \kappa) & \alpha & 0 & 0 & \beta \frac{N}{N_v} \\
0 & \kappa & - (\lambda + \gamma) & 0 & 0 & 0 \\
0 & 0 & \beta_v \frac{N_v}{N} & - (\lambda + \mu) & 0 & 0 \\
0 & 0 &  \beta_v \frac{N_v}{N} & 0 & -  (\lambda + \mu + \eta) & 0  \\
0 & 0 & 0 & 0 & \eta & - (\lambda + \mu) 
\end{bmatrix}
= 0.
\]

We can reduce this equation to a product of two factors and a fourth degree polynomial equation
\[
\lambda(\lambda + \mu) 
det \begin{bmatrix}
-(\lambda + \kappa) & \alpha  & 0 & \beta\frac{N}{N_v} \\
\kappa & - (\lambda + \gamma) & 0 & 0 \\
0 & \beta_v \frac{N_v}{N} & - (\lambda + \mu + \eta) & 0 \\
0 & 0 & \eta & - (\lambda + \mu) 
\end{bmatrix}
= 0.
\]

The initial exponential growth rate is the largest root of this fourth degree equation, which reduces to
\begin{equation} \label{grow2}
g(\lambda) = (\lambda + \kappa)(\lambda + \gamma)(\lambda + \mu + \eta)(\lambda + \mu) - \beta \beta_v \kappa \eta 
                       - \kappa \alpha (\lambda + \mu)(\lambda + \eta + \mu)  = 0.
\end{equation} 
The largest root of this equation is the initial exponential growth rate, and this may be measured experimentally. If the measured value is $\rho$, then from \eqref{grow2} we obtain
\begin{equation} \label{rho}
(\rho + \kappa)(\rho + \gamma)(\rho + \mu + \eta)(\rho + \mu) - \beta \beta_v \kappa \eta 
                       - \kappa \alpha (\rho + \mu)(\rho + \eta + \mu)  = 0.
\end{equation}
From \eqref{rho} we can see that $\rho = 0$ corresponds to $\mathcal{R}_0 = 1$, confirming that our calculated value of 
$\mathcal{R}_0$ has the proper threshold behavior.

The equation \eqref{rho} determines the value of $\beta \beta_v$, and we may then calculate  $\mathcal{R}_0$, provided we know the value of $\alpha$. However, this presents a major problem. In \cite{Getal16} it is suggested that the contribution of sexual disease transmission is small, based on estimates of sexual activity and the probability of disease transmission. Since the probability of sexual transmission of a disease depends strongly on the particular disease, this estimate is quite uncertain. Estimates based on a possible imbalance between male and female disease prevalence are also quite dubious. Most Zika cases are asymptomatic or quite light but the risks of serious birth defects means that diagnosis of Zika is much more important to women than to men. If there are more female than male cases, it is not possible to distinguish between additional cases caused by sexual contact and cases identified by higher diagnosis rates. To the best of our knowledge, there is not yet a satisfactory resolution of this problem. 

What would be required would be another quantity which can be determined experimentally and  can be expressed in terms of the model parameters. After an epidemic has passed, it might be possible to estimate the final size of the epidemic, and then it would be possible to choose values of $\alpha$ and $\beta \beta_v$ satisfying \eqref{rho} such that simulations of the model \eqref{model2} give the observed final size.

\section{Discussion}

We have examined an SEIR/SEI vector-transmission
epidemic model that may be applied to dengue fever, chikungunya virus,
and Zika virus outbreaks. In addition, we have examined a novel model for Zika virus outbreaks that 
also includes direct sexual transmission. For both models, we have obtained expressions for the reproduction number and ways of estimating the initial exponential growth rate, so that the reproduction number may be estimated from parameters that can be estimated. There are no exact analytic solution for  final size relation, 
but numerical simulations can be used for predictions.

While Zika and chikungunya virus have only one serotype in humans, dengue virus
has four serotypes, with potential cross-immunity between strains.  
The models we have examined do not include the effect of multiple serotypes.

In spite of these shortcomings, our models can be used to simulate the effects of different control strategies, including mosquito control, reduction of contact with mosquitoes, and avoidance of sexual contact (for Zika).  It might be worthwhile to formulate and analyze a Zika model with hosts divided into males and females, but at present it is unlikely that such a more detailed model would provide better information about the course of the epidemic.

It should be pointed out that control measures which decrease the mosquito population will decrease the rate of bite of humans because of the balance relation \eqref{bal}, and will thus decrease the reproduction number. However, measures that protect some humans from being bitten will only redistribute bites to other humans and thus introduce heterogeneity of bites and require an adjustment to the model to include two classes of humans with different rates of being bitten. This suggests that control strategies aimed at decreasing the number of mosquitoes may be much more effective than measures protecting against being bitten. However, it is possible that if the supply of human victims is insufficient, mosquitoes may shift some of their bites to animals. This would destroy the balance equation, and would lead to a need to model vector disease transmission if there is more than one host species.
Extending the model to more than one class of hosts would allow studies of comparison of disease spread in different populations, such as large cities versus small villages, or rich and poor populations with different degrees of protection mechanisms.

From a mathematical perspective, we point out that for vector disease transmission models there are ambiguities in the calculation of the reproduction number using the next generation matrix approach. Different choices in the next generation matrix approach may give different expressions for the basic reproduction number. Different choices are made for the basic reproduction number in different studies, and studies do not always indicate clearly what choices have been made. This may lead to misinterpretations. 

For example, the usual next generation matrix approach gives a square root in the reproduction number because it views the transition from host to vector to host as two generations. It is common, but by no means universal, to remove this square root from the reproduction number. A model with both direct and vector disease transmission such as \eqref{model2} indicates that removal of the square root is logically sound, and that care is needed in the calculation of the reproduction number for a vector transmission model.

\section{Acknowledgements}

The authors are grateful for the funding support received from the following sources:

\noindent FB Natural Sciences and Engineering Research Council (Canada) \\
     Grant OGPIN 203901 - 99 \\
\noindent CCC, AM, ST National Science Foundation (DMS - 126334, DUE - 1101782) \\
            National Security Agency (H98230 - 14 -1 -0157) \\
            Office of the President of Arizona State University \\
            Office of the Provost of Arizona State University

\end{document}